# Molecular dynamics simulation of W Silicon Emitting Centers formation by Ga ion implantation

Christos Gennetidis and Patrice Chantrenne
*Univ. Lyon, INSA-Lyon, Université Claude Bernard Lyon 1, CNRS,
MATEIS, UMR 5510,69621 Villeurbanne, France*

Thomas Wood
*Univ Lyon, Ecole Centrale de Lyon, CNRS, INSA Lyon, Université Claude Bernard Lyon 1, CPE Lyon,
CNRS, INL, UMR5270, 69130 Ecully, France*

Silicon Emitting Centers (SEC) constitute promising candidates for quantum telecommunication technologies. Their operation depends on the fabrication of light emitting defect centers such as the tri-interstitial Si complex, the W-Center. In this paper the formation of Si tri-interstitial clusters after Ga ion beam bombardment on pure silicon substrates and a subsequent annealing stage is investigated using molecular dynamics (MD) simulations. This study aims to understand the dynamic formation process of W centers after Ga implantation and annealing in order to assist the focused ion beam and annealing experimental systems. A new tri-interstitial cluster identification method is proposed which considers the configuration of the clusters in the Si lattice in order to identify the defects which will act as candidates for the W center. This method successfully identifies W center defect candidates in an ideal system. The number of tri-interstitial clusters increases and spread deeper into the Si for higher energies and their probability of generation increases until a limiting Ga dose. Furthermore, annealing can eliminate a lot of the unwanted defects maintaining at the same time the number of the tri-interstitial clusters, leading to isolated clusters with less distorted local environment.

## I. INTRODUCTION

Single photon emitter technology has drawn the attention of researchers in recent years due to its application in various fields, including quantum telecommunications. From a plethora of candidate materials capable of hosting single photon emitters, silicon constitutes a material with a wide range of advantages. It may host a large number of point defect-based emitters such as the G and W centers, which incorporate energy levels inside the bandgap and emit near the telecommunication wavelength bands[1]. Furthermore, Silicon is a well-known material which can be integrated to Si microelectronics devices. This, combined with its natural abundance and high purity, makes Silicon the most promising candidate material for quantum applications [2].

The W Silicon Emitting Center (SEC) is a promising defect which has not been investigated extensively yet. It consists of a Si self tri-interstitial cluster which demonstrates a trigonal $C_{3v}$ symmetry with its $C_3$ axis parallel to the ⟨111⟩ crystallographic directions, as was revealed by uniaxial stress measurements carried out by Davies et al. [3]. The precise structure of the W center is still under debate. The I3-I structure consists of three interstitials in an equilateral triangle in a {111} plane bridging three adjacent bond centered sites [4], whereas the I3-V structure is a wider configuration of the I3-I [5, 6]. Both exhibit the $C_{3v}$ symmetry and Local Vibrational Mode energy at 70meV [6, 7]. However, recently Baron et al. [7] showed by Density Functional Theory simulations that the I3-V defect can account for an optically active W center whereas the I3-I cannot.

One way to fabricate these defects is the employment of a Focused Ion Beam (FIB) implantation method which can incorporate the desired defects inside the bulk material with defined depth and high lateral resolution. FIB has been employed to fabricate the Silicon vacancy center in diamond [8] and G and W centers in Silicon with the use of a Si beam [9]. Whether FIB implantation or wafer-scale implantations are used, subsequent annealing stages have been shown to annihilate a lot of the damage created during the implantation and isolate the desired defects which will act as potential photon emitting centers [3, 7–13]. A popular ion beam element in FIB systems is Gallium, due to its low vaporization temperature, which, due to its high mass, can lead to a large amount of damage in the implanted substrate. It is widely used for transmission electron microscopy sample preparations [14–16]. In this applicative domain, molecular dynamics (MD) simulations have been employed to investigate the FIB milling process under different experimental conditions to study the amorphization of the sample [14, 16– 19]. Furthermore, Xiao et al. [20] studied using MD simulations the damage formation during the implantation of Ga into Si, as well as the damage evolution after high temperature annealing, but they did not focus on the formation of any specific kind of defects.

MD simulations is a useful tool to help elucidate the dynamic formation process of different defect centers during the FIB implantation and subsequent annealing stages. This is its main advantage over other simulation methods such as the Binary Collision Approximation (BCA) method of SRIM software [21]. SRIM is parameterized with experimental implantation data, so it can calculate with good accuracy the mean stopping depth of the implanted ions. On the other hand, because it treats the lattice as amorphous, it cannot investigate dynamic processes at different temperatures such as the annealing as well as identify complex defect structures in a crystal lattice such as a tri-interstitial cluster, which is the origin of the W SEC.

Several studies employed MD simulations to investigate different potential single photon emitting defects after implantation, especially on SiC [22–24] and diamond [25–27]. Fan et al. [23] used H ion implantation whereas in ref. [22] they used dual ion implantation of He and Si to investigate the formation and evolution of Si vacancy defect centers under different annealing conditions. Fu et al. [26] implanted Si into diamond and used annealing at different temperatures in order to create Si vacancy centers. Regarding the Si system, Aboy et al. [28] used MD simulations to investigate the formation and evolution of small Si self-interstitial clusters after placing a number of interstitial Si atoms close to each other and annealing the system for 25ns at 1200K. Using the Tersoff 3 interatomic potential [29] to account for the Si-Si interactions they found the formation of the I3-V defect but they didn't find the I3-I defect. So far there is neither an experimental nor computational work investigating the formation process of W centers during Ga ion implantation on Silicon.

In this work, the formation of W center candidates during the implantation of Ga atoms into Si and the subsequent annealing stage was investigated by the use of MD simulations. A new W center candidate identification algorithm was developed which takes into account the position and symmetry of the defect which, according to the literature [4–7], constitutes a W center. Different implantation energies and doses compatible with the FIB process as well as different annealing temperatures and times were studied. This investigation aims to understand the formation and evolution process of W centers under the different experimental conditions.

## II. METHODS

The molecular dynamics simulations were carried out using the LAMMPS [30] simulation software. The interatomic interactions were described by the Tersoff [31] potential smoothly connected with the universal ZBL (Ziegler Biersack Littmark) screened nuclear repulsion potential to account for the high-energy collisions between the atoms. During the Ga-Si interactions only the nuclear stopping (elastic collisions) from the ballistic collisions was taken into account. The electronic stopping from inelastic collisions of the atoms with the electrons is considered to have minor impact on slow moving heavy atoms such as Ga [32]. The potential parameters employed in this work for Si and Ga can be seen in references [31, 33], where the mixing rules [31] have been used to determine the parameters for Ga-Si interactions.

The simulation model consists of a Si atoms box and Ga atoms randomly initialized 1nm above the (001) plane of Si in order to be implanted to it with a zero incidence angle to the (001) plane's normal. Four Ga beam energies were investigated (0.5keV, 1keV, 2keV and 5keV) which are consistent with the low energy operation of FIB. Two system sizes were used: a small one (217800 atoms) for the 0.5keV, 1keV and 2keV and a bigger one (579200 atoms) for the 5keV. Fig 1 depicts the small size implantation model where the bottom atoms (brown) are fixed and the rest of the atoms (yellow) are free to move under Newtonian interactions with periodic boundary conditions across x and y directions. Each Ga atom is implanted one at a time in a square region of 10 × 10$nm$ and the system is thermalized at 300K between each implantation. The implantation process was divided into two parts. In the first, each Ga implantation starts with a timestep of 0.02fs and NVE ensemble to follow the high energy atom movements and allow the cascade to form naturally. The use of this timestep, as well as with all the different timesteps employed in this study, ensures the conservation of the total energy of the system. The first part lasts 0.4ps which is enough for the implanted Ga atom to transfer its energy to the Si lattice. This is within the 0.1-1ps time scale of the ballistic collision processes [34], where the high energy collisions between the atoms occur. In the second part the NVT thermostat was used on the whole system with a timestep of 0.5fs for 20ps to slowly cool down the system back to 300K.

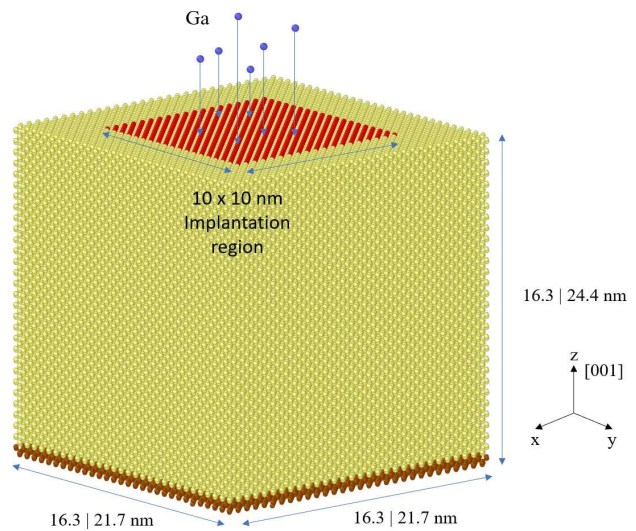

FIG. 1. MD model of implantation. Blue atoms correspond to the implanted Ga atoms, brown atoms to the bottom fixed layer, yellow atoms to the Newtonian region and the red atoms to the surface atoms where the implantation occurs. The different lengths at the sides indicate the side lengths of the different models employed in this study.

After the implantation process the systems were annealed at different temperatures and for different times. Before each annealing the implanted structure was equilibrated at 300K and zero pressure (NPT) for 50ps. Then the systems were heated with a heating rate of 0.02K/fs to the desired annealing temperature, with a timestep of 0.2fs. After the annealing the system was quenched at the same rate to 300K where it was thermalized for 50ps under NPT followed by a 100ps stage under the NVE ensemble. The average number of defects was calculated using the methods described below, every 5ps at this final NVE stage.

The Wigner-Seitz (WS) analysis cannot properly locate all the tri-interstitial clusters, especially those which, according to the literature [4–7] are more probable to act as an optically active W center. Thus, a new identification method was developed using Python programming alongside the OVITO WS analysis [35] to identify the best candidates for the W center. Firstly, the WS analysis was employed to select clusters that have atoms which are recognized as interstitials. Then all the clusters in between the {111} planes, which have a maximum distance between their atoms of $3.5\text{Å}$, as described later, were located. This distance was found to be sufficient to identify the desired clusters (section III.A). Afterwards, the ones for which the normal to the plane containing the atoms formed an angle of less than 10 degrees with the ⟨111⟩ directions were selected. This way, the best tri-interstitial candidates for the W center were selected which will be referred simply as tri-interstitial clusters. Furthermore, this algorithm has the capability to identify from these candidates the ones with the appropriate trigonal symmetry by calculating the interior angles of the tri-interstitial clusters. The symmetric clusters were selected by finding the ones for which the three interior angles did not differ by more than 10 degrees. These clusters will be referred as symmetric tri-interstitial clusters.

## III. RESULTS AND DISCUSSION

### A. Ideal W centers

The formation by the employed potential of a tri-interstitial cluster which can account for a W center according to the literature Ref. [4–7] was investigated. In a small Si box of $9 \times 9 \times 9$ lattice constants and 5832 atoms with periodic boundary conditions across each direction a distorted tri-interstitial cluster was placed in between the {111} planes. Afterwards, the structure was minimized to find its ground state, and two that resemble the I3-I and I3-V structures of the literature were found (Fig 2). These structures were annealed at 300K to investigate their stability. The I3-I defect remained stable fluctuating around its initial position due to the thermal movements, with the tri-interstitial atoms having a distance of around $2.4\text{Å}$. The I3-V defect exhibits a small rotation and the distance between the tri-interstitial atoms falls from $3.8\text{Å}$ at 0K to $3.4\text{Å}$ at 300K. In any case they appear to preserve their trigonal symmetry and positioning at 300K, which validates the use of the employed interatomic potential. Furthermore, the developed algorithm can successfully locate these defects at 300K and 1K by setting the maximum thresholds for the distance between the atoms at $3.5\text{Å}$ and the angles at 10 degrees.

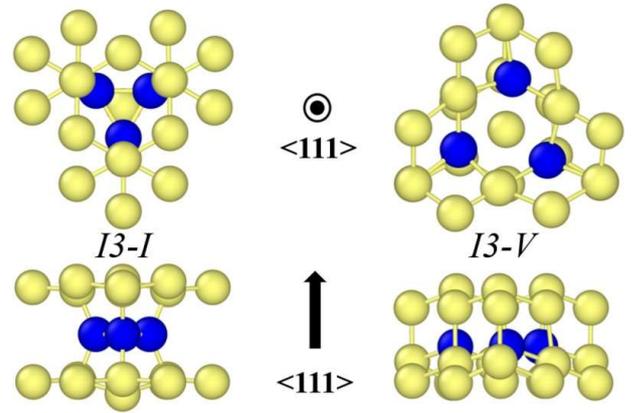

FIG. 2. W center defect structure candidates, produced by minimization of a distorted tri-interstitial cluster placed in between the {111} planes using the Tersoff potential. W center with the I3-I (left) and I3-V (right) structure at 0K. The blue balls correspond to the tri-interstitial defect atoms.

### B. Implantation energy and dose

In this section the effect of the implantation energy and dose on the formation of defects is presented. In Fig 3 the mean implantation depth of the Ga atoms and the mean formation depth of WS interstitials, WS vacancies as well as the tri-interstitial clusters identified from the developed method is shown for different energies and a dose of $50 \times 10^{12} ions/cm^2$ for one simulation of 50 successively implanted Ga atoms. The same parameters were employed for SRIM simulations and the ion ranges of Ga atoms are plotted alongside the ones from MD simulations. The two simulation methods give similar results for the Ga mean implantation depth even though no direct comparison can be made between the two methods because of the crystal and amorphous structures of MD and SRIM respectively. The MD simulation with 5keV beam energy exhibits channeling effects, which can be eliminated by a small beam angle of $6^o$. The WS defects as well as the tri-interstitial clusters appear to have the same mean depth at each energy which is approximately 2nm below that of Ga. Their average depth distributions from ten different simulations for the different implantation energies are shown in Fig 4. The tri-interstitial clusters are formed near the highly defected region of interstitials and vacancies for the low energies of 0.5keV and 1keV, whereas they spread deeper into the sample for the higher energies and especially for the 5keV, where some tri-interstitial clusters are formed far from the highly defected area (below $100\text{Å}$). The high Ga energies allows them to transfer their energy deeper into the sample and create tri-interstitial clusters in a less distorted environment.

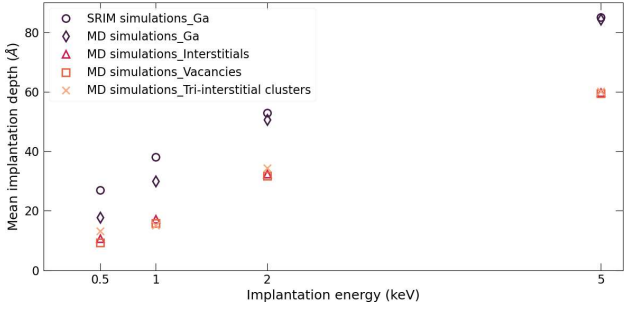

FIG. 3. Ga and defects mean implantation depth as a function of implantation energy for MD and SRIM simulations.

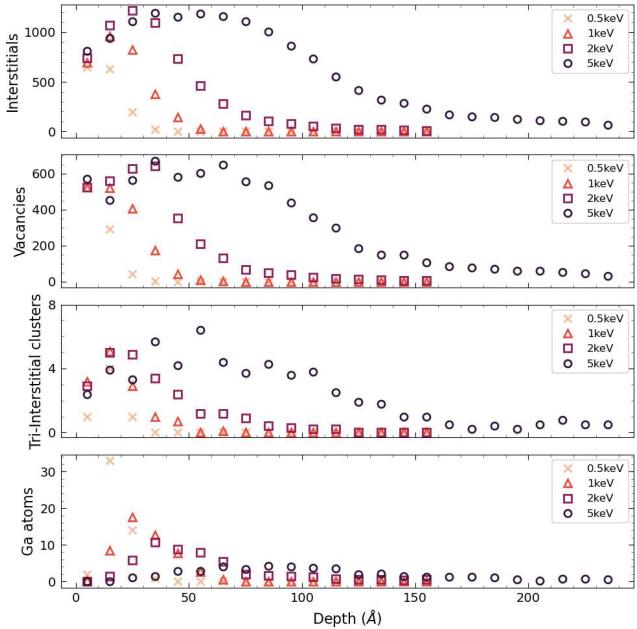

FIG. 4. Interstitials (a), vacancies (b), tri-Interstitial clusters (c) and implanted Ga atoms (d) average depth distributions (ten simulations), for different implantation energies with a dose of $50 \times 10^{12} ions/cm^2$.

The 5keV implantation energy yields the most tri-interstitial clusters deeper into the Si, thus a Ga dose investigation was carried out for this energy. In Fig 5 the number of tri-interstitial clusters with the appropriate positioning and orientation (Fig 5 top) as well as the trigonal symmetric ones (Fig 5 bottom) are presented for different implantation doses. The orange circles for the same dose represent the number of the clusters produced after a given number of implantation events for multiple simulations and the black dots the average number over all simulations.

The number of non-symmetric tri-interstitials appears to increase linearly with the dose up to $60 \times$

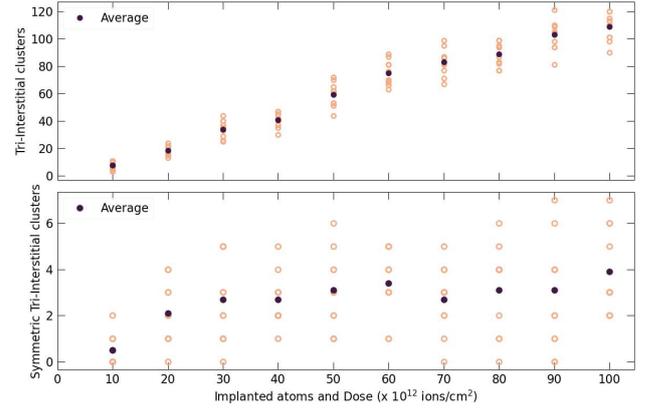

FIG. 5. Non-symmetric tri-Interstitial clusters (top) and symmetric tri-interstitial clusters (bottom) as a function of the implantation dose (and implanted atoms) for 5keV implantation energy. The orange circles represent the number of the clusters produced after a given number of implantation events for multiple simulations and the black dots the average number over all simulations.

$10^{12} ions/cm^2$, after which it increases more slowly. The trigonal symmetric tri-interstitials reach an approximately stable average after a dose of $30 \times 10^{12} ions/cm^2$ but their number is not sufficient to extract meaningful statistical results. Thus, for the following analysis only the non-symmetric clusters will be used. Moreover, it is possible that with the appropriate thermal energy a transition can occur from the non-symmetric to the symmetric tri-interstitial cluster. One possible mechanism for this transition may be the translation and rotation during the diffusion of the clusters which requires relatively small energies [36]. Because of this, the non-symmetric tri-interstitials clusters can be considered W center defect candidates.

### C. Annealing temperature

The annealing temperature is an important parameter [3, 7–13] for the formation of silicon emission centers. The MD simulations cannot be used to investigate the full annealing process in terms of time but the trend the defect populations follow during short annealing times can be investigated. Thus, annealing was carried out on the system of 5keV implantation energy and a dose of $50 \times 10^{12} ions/cm^2$ for 0.5ns at different temperatures in the range of 400K-1400K. The results for the tri-interstitial clusters can be seen in the top of Fig 6 for the different annealing temperatures. The number of clusters fluctuates widely especially at low annealing temperatures. This happens because a lot of these clusters are located in the highly defected area and transition from tri-interstitial clusters to other kind of defects due to their distorted environment. The mean number of clusters decreases slowly till the 600K and then it decreases more rapidly up to 1200K. From there the number of clusters increases for the higher annealing temperatures. This behavior at high temperature may be explained by the high kinetic energy of the atoms which leads to faster defect diffusion for the same annealing time and eventually to the formation of more tri-interstitial clusters.

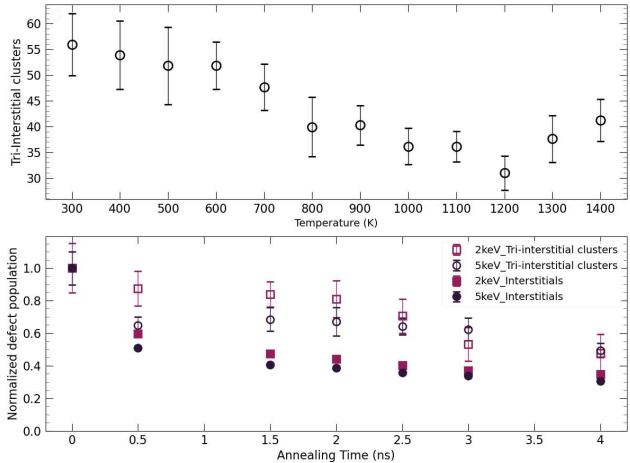

FIG. 6. Top: Tri-Interstitial clusters after 0.5ns of annealing at different temperatures for the 5keV implantation energy and a dose of $50 \times 10^{12} ions/cm^2$. Bottom: Normalized number of tri-interstitial clusters and the overall interstitials after different annealing times at 1400K for different implantation energies and a dose of $50 \times 10^{12} ions/cm^2$.

### D. Annealing time

In this section the annealing time at 1400K for each implantation energy and a dose of $50 \times 10^{12} ions/cm^2$ was investigated. This temperature was selected because it will lead to faster defect evolution and eventually better annihilation of the lattice damage. The systems were annealed for 0.5, 1.5, 2, 2.5, 3 and 4ns and the normalized average number of interstitials which constitutes unwanted damage to the crystal structure and tri- interstitial clusters can be seen at the bottom of Fig 6 for 2keV and 5keV. For the smaller energies the number of tri-interstitial clusters is small and their annihilation rate is similar to that of the interstitials. On the contrary for 2keV and 5keV the total number of interstitials is reduced with the annealing time falling below 50% after 1.5ns of annealing whereas the number tri-interstitial clusters fall below 50% after 4ns of annealing. Furthermore, the tri-interstitial clusters appear to stabilize especially for the 5keV in the region 0.5ns-3ns and experiencing even a small increase at 1.5ns. This shows that the rate of annihilation of the two defects is different with the number of interstitials reducing quicker than the one of the tri-interstitial clusters. As a consequence, this is an indication that the annealing can eliminate a lot of the unwanted defects, retaining at the same time the popu lation of the tri-interstitial clusters stable. A visual representation of this can be seen in the defect evolution snapshots of Fig 7. Apart from the increase of the tri-interstitial clusters to interstitial ratio, it is obvious that a lot of tri-interstitial clusters with less distorted local environment (Fig 7b and Fig 7c) are formed. Before the annealing (Fig 7a) the highly defected environment may have detrimental effect to the stability of the potential W centers. Thus, the annealing can further improve the stability and isolation of the tri-interstitial clusters by having a less distorted local environment.

## IV. CONCLUSIONS

In this work molecular dynamics simulations were employed to investigate the formation of W centers under different experimental FIB parameters for the Ga implantation into Si as well as the subsequent annealing stage. A new identification method was developed to locate potential W center candidates. It can locate tri-interstitial clusters positioned in between the {111} planes with the appropriate ⟨111⟩ orientation as well as symmetric tri-interstitial clusters which furthermore have trigonal symmetry. The low implantation energies do not yield many tri-interstitial clusters, which are furthermore concentrated in the highly defected areas. On the other hand, the 5keV implantation energy produces more tri-interstitial clusters which are also spread deeper into the Si where the damage is not so severe. The tri-interstitial cluster population increases with the Ga dose with a decreasing rate whereas the symmetric tri-interstitial clusters appear to stabilize after a dose of $30 \times 10^{12} ions/cm^2$. The annealing for the 2keV and 5keV implantation energies showed that the overall interstitial populations decreases more rapidly than the tri-interstitial clusters with increasing annealing time. For intermediate annealing times the 5keV tri-interstitial clusters appear to stabilize and even increase at some point. This is an indication that the annealing can annihilate a lot of the unwanted defects and at the same time maintain the tri-interstitial clusters which can acquire a better, less distorted local environment. Eventually this could lead to more stable and isolated tri-interstitial clusters and, potentially, optically active W centers. Symmetric tri-interstitial clusters can be found after annealing (top of Fig 7) but their numbers are not sufficient to extract statistical results. This study hopes to further the understanding of the dynamical formation and evolution of W centers after FIB implantation and the subsequent annealing stage and guide future experimental and theoretical work.

## ACKNOWLEDGMENTS

We acknowledge financial assistance from the INSA de Lyon 'Bonus Qualit´e Recherche' grant scheme, 2021.

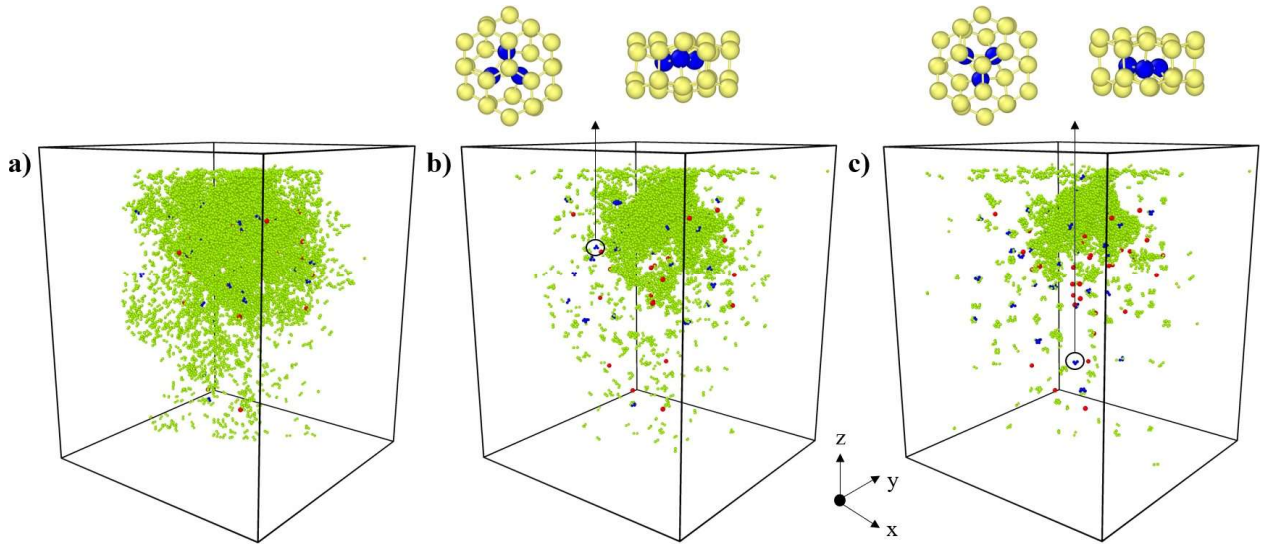

FIG. 7. Snapshots of the defects in the 5keV implantation system with a dose of $50 \times 10^{12} ions/cm^2$ a) before annealing, b) after 0.5ns annealing and c) 3ns annealing at 1400K. The Ga atoms are represented with red, the interstitials with green and the tri-interstitial clusters with blue. In the black circles and above the simulation box are some of the symmetric tri-interstitials identified by the developed algorithm observed following the ⟨111⟩ (left image) and ⟨11 − 2⟩ (right image) directions.